\documentclass[twocolumn,showpacs,prb,amsmath]{revtex4}
\usepackage{bm}
\usepackage{epsfig}
\usepackage{ulem}
\newcommand{\nix}[1]{}

\begin{document}

\title{Tuning of structure inversion asymmetry by the $\delta$-doping position in (001)-grown GaAs quantum wells}
\author{ V.~Lechner,$^1$ L.E.~Golub,$^2$ P.~Olbrich,$^1$ S.~Stachel,$^1$
D.~Schuh,$^1$ W.~Wegscheider,$^1$ V.V.~Bel'kov,$^{1,2}$ and S.D.~Ganichev$^{1}$
}
\affiliation{$^1$Terahertz Center, University of Regensburg,
93040 Regensburg, Germany}
\affiliation{$^2$Ioffe Physico-Technical Institute, Russian
Academy of Sciences, 194021 St.~Petersburg, Russia}

\begin{abstract}
Structure and bulk inversion asymmetry in doped (001)-grown GaAs quantum wells is investigated by
applying the magnetic field induced photogalvanic effect. We demonstrate that the structure
inversion asymmetry (SIA) can be tailored by variation of the delta-doping layer position. 
Symmetrically-doped structures exhibit a substantial
SIA due to impurity segregation during the growth process.
Tuning the SIA by the delta-doping position we grow samples 
with almost equal degrees of structure and bulk inversion asymmetry.
\end{abstract}

\maketitle

The generation, manipulation and detection of spin polarized electrons
in low dimensional semiconductors are at the heart of spintronics, see
e.g. Refs.~\onlinecite{Winkler07,Fabian07,Dyakonov08}.
A versatile tool to achieve these goals provides spin-orbit coupling
which in quantum wells (QWs) based on  III-V
semiconductors removes the spin degeneracy of the energy bands.
The spin-splitting allows one to control the spin polarization by
the electric field, determines the spin
relaxation rate and can be utilized for all-electric spin
injection.
The lifting of the spin degeneracy is caused by spin-orbit
interaction described by    linear in electron wavevector
${\bm k}$ terms in the effective Hamiltonian
$H_{SO}=\sum\beta_{lm}\sigma_lk_m$, where ${\bm \beta}$ is a second
rank pseudo-tensor and $\sigma_l$ is the Pauli matrix.
The microscopic origin of these terms
is the structure inversion asymmetry 
and bulk inversion asymmetry  (BIA) which lead to Rashba and Dresselhaus
spin-orbit terms in $H_{SO}$, respectively.~\cite{Winkler07,Fabian07,Dyakonov08}
The strength of the BIA  spin-splitting depends on the QW width,
temperature, and electron density. SIA
stems from the inversion asymmetry of the confining
potential.
The widely applied method of the variation of the SIA related effects
is the application of a gate voltage.
Recently we demonstrated that in doped (110)-oriented GaAs QWs the proper choice of
the $\delta$-doping layer position permits the growth of structures with controllable SIA.~\cite{condmat110}

Here we investigate (001)-grown GaAs QWs
and demonstrate that
SIA is substantially affected by the segregation during the structure growth.
This is in contrast to (110)-oriented QWs
where the segregation is suppressed.~\cite{condmat110}
Our experiments explore the role of segregation and allowed us
to determine the growth conditions of structures with predetermined SIA.
By that, we prepared
the QWs with almost equal Rashba and Dresselhaus spin-splittings.
Such structures  can be applied for the development of a non-ballistic spin-field effect
transistor,~\cite{Schliemann03p146801} creation of persistent spin helix~\cite{helix} and are
characterized by a drastic increase of spin relaxation times.~\cite{Averkiev08p114002}

The method to explore the BIA and SIA degrees is
based on the investigation of anisotropy of the magneto-photogalvanic effect (MPGE).
The MPGE is a photocurrent generation driven
by absorption of radiation in
QWs in the presence of a magnetic field.~\cite{Belkov08}
The effect is caused by the lack of a space inversion symmetry, thus
the strength and the direction of the photocurrent is determined by
the interplay of  SIA and BIA.
The MPGE current density
is given by~\cite{Belkov08}
\begin{equation} \label{phen0}
j_\alpha = \sum_{\beta\mu\nu}
\phi_{\alpha\beta\mu\nu}\:B_\beta\:(E_\mu
E^*_\nu + E_\nu  E^*_\mu)/2\:,
\end{equation}
where $\bm{\phi}$ is a fourth rank pseudo-tensor being symmetric in the indices
$\mu$ and $\nu$, $B_{\beta}$ are components of the magnetic field $\bm{B}$,
and $E_\mu$ are the components of the radiation electric field $\bm E$.
Excitation of (001)-grown QWs with unpolarized radiation at normal incidence in
the presence of an external
in-plane magnetic field  provides a straight forward method
to obtain the relative strengths and sign of SIA and BIA.
In this case Eq.~\eqref{phen0} reduces
to
%
$   j_l=\sum_m \gamma_{lm} B_m |\bm E|^2$,
where $\bm{\gamma}$
is a second rank pseudo-tensor.
It is seen
that both MPGE and the spin-splitting are characterized by the same anisotropy in space
because they are described by equivalent second rank pseudo-tensors $\bm \gamma$ and $\bm \beta$ whose irreducible
components differ by a scalar factor only.

In analogy to the band spin-splitting
the photocurrent can be decomposed into SIA and BIA contributions.
In particular for an in-plane magnetic field applied
along a cubic axis, e.g. $y$-direction, we obtain
for longitudinal, $j_y \parallel [010]$, and transverse,  $j_x \parallel [100]$,
photocurrents
\begin{equation}
\label{MPGE_second}
j_{x} = \gamma^{\rm SIA}\:B_y\,|\bm E|^2,
\quad
j_{y} =  \gamma^{\rm BIA}\:B_y\,|\bm E|^2\:.
\end{equation}
Here $\gamma^{\rm SIA}$ and $\gamma^{\rm BIA}$ are components of 
the tensor ${\bm \gamma}$ due to SIA and BIA, respectively.
Taking the ratio $j_x/j_y$
cancels the scalar factor which contains all microscopic
details
and yields directly the ratio of SIA to BIA.
Moreover separate analysis of photocurrents  given by Eq.~\eqref{MPGE_second}
reveals changes of the strength and the sign of SIA and BIA upon a variation
of external parameters.

\begin{figure}
\centerline{\epsfxsize 70mm \epsfbox{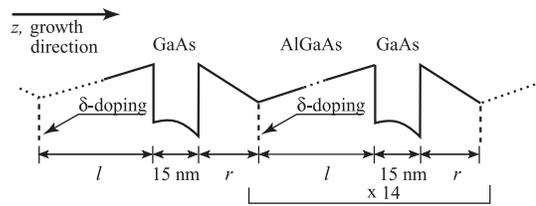}}
\caption{Band profile of QWs and doping position.}
\label{fig1}
\end{figure}

\begin{figure}
\centerline{\epsfxsize 70mm \epsfbox{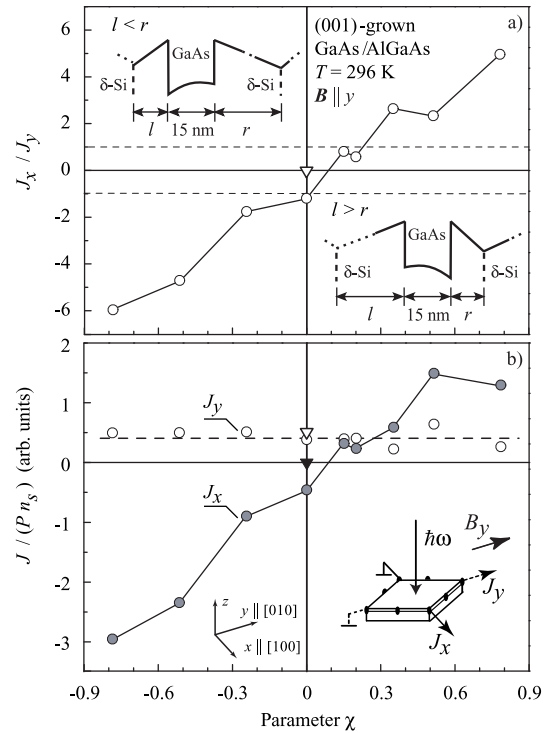}}
\caption{a) The ratio of the SIA and BIA contributions to the MPGE, $J_x/Jy$, as a function of $\chi$.
Triangle shows the result for sample~5LT grown at $T_\delta = 490^\circ$C, circles demonstrate the data for all other samples 
grown at $T_\delta ~\approx 630^\circ$C. Insets show the QW profile and the  doping positions for $l < r$
and for $l>r$.
b) Dependence $J/P n_s$ on the parameter $\chi$. The photocurrents are measured
along and normal to $\bm B || y$.
Full and open symbols show $J_x$ and $J_y$, respectively (triangles are the data for sample~5LT).
Inset shows the experimental geometry.
}
\label{fig2}
\end{figure}

We investigated (001)-oriented
Si-$\delta$-doped $n$-type  GaAs$/$Al$_{0.3}$Ga$_{0.7}$As  structures grown by
molecular-beam epitaxy at temperatures over 600$^\circ$C.
Temperatures $T_\delta$ during the growth of the $\delta$-doping layer and subsequent 
layers, which control the segregation, mobilities $\mu$ and carrier densities $n_s$
measured at room temperature are given in Table~\ref{table1}.
Figure~\ref{fig1} sketches the conduction band  edge
together with the corresponding $\delta$-doping position.
All QWs have the same width of 15~nm but differ essentially  in their doping profile:
In all structures besides sample 4 and 5LT,
the doping layers are asymmetrically
shifted off the barrier center either to the left  or to
the right. The impurities' Coulomb field
yields an asymmetric potential profile inside the QWs.
To describe the degree of asymmetry we introduce the parameter $\chi = (l-r) / (l+r)$, where
$l$ and $r$ are the  spacer layer thicknesses between QW and $\delta$-layers
(Fig.~\ref{fig1} and Table~\ref{samples}).
Samples 4 and 5LT contain a Si-$\delta$-sheet, placed in the middle of each barrier
between adjacent  QWs.
All samples were square shaped
with the sample edges  of 5~mm length oriented along
$[1{\bar 1}0]$ and $[110]$ crystallographic axes.
In order to measure photocurrents, ohmic contacts were alloyed on the sample corners and
in the middle of each  sample side allowing to probe the photocurrent in
different directions, as displayed in Fig.~\ref{fig2}.
An external in-plane magnetic field $B = \pm 1$~T is applied
along either  $x \parallel [1 0 0]$ or $y \parallel [0 1 0]$.
\begin{table}
\caption{\label{table1}Parameters of samples. Carrier density $n_s$ (per QW-layer) and mobility
$\mu$ are given for room temperature.}
\begin{tabular}{cccccccc}
\hline %

 sample      & spacer  & spacer &\, $\chi=\frac{l-r}{l+r}\,\,$ &  $n_s$  &  $\mu$  &  $T_\delta$ \\
       & \,\, $l$, nm   & \,\,$r$, nm &                        & $10^{11}$$\frac{\rm{1}}{\rm{cm}^2}$    & \, $10^{3}$$\frac{\rm{cm}^2}{\rm{Vs}}$  &  $^\circ$C \\[0,07cm]
\hline
  1 &  20   & 165  & -0.78 & 1.2  & 7.2   &  615 \\ \hline
  2 &  45   & 140  & -0.51 & 1.6  & 7.7   &  631 \\ \hline
  3 &  70   & 115  & -0.24 & 1.4  & 8.0   &  631 \\ \hline
  4 &  92.5 & 92.5 & 0     & 1.4  & 7.7   &  631 \\ \hline
  5LT &  92.5 & 92.5 & 0     & 1.0  & 7.9   &  490 \\ \hline
  6 & 106.5 & 78.5 & 0.15  &  1.4 &  8.1  &  626 \\ \hline
  7 &  111  & 74   &  0.20  &  1.4 &  8.1  &  632 \\ \hline
  8 & 125  & 60   &  0.35 & 1.4  &  7.7  &  629 \\ \hline
  9 &  140 & 45   & 0.51  & 1.4  & 7.9   &  630 \\ \hline
  10 &  165 & 20   & 0.78  & 1.5  & 8.0   &  625 \\
  \hline
 \end{tabular}%
 \label{samples}%
\end{table}

The MPGE is measured at room temperature
by exciting the samples with unpolarized
terahertz radiation under normal incidence, as sketched in the inset of Fig.~\ref{fig2}b. The pulsed
radiation of power $P\approx 5$~kW is obtained
applying an optically pumped pulsed NH$_3$ molecular laser.~\cite{GanichevPrettl}
The wavelength of $280\,\mu$m was chosen to cause only
free carrier absorption of the radiation.
To obtain
unpolarized light
we used a brass cone of 95~mm length with the angle of about 4$^\circ$ which depolarizes
the laser radiation as a result of multiple reflection in the cone.

Irradiating samples
with unpolarized  radiation
we observed a photocurrent
signal linearly increasing with rising
magnetic field strength and changing
the sign upon inversion of the
magnetic field direction from $B_+ > 0$ to $B_- < 0$.
For convenience in the discussion below we evaluate the data after
$J = \left[J(B_{+})-J(B_{-})\right]/2$ which yields solely MPGE contribution.
The signal
is detected in both longitudinal and transverse
directions with respect to the magnetic field orientation.

Figure~\ref{fig2}a shows the ratio  between $J_x$
and $J_y$ photocurrents  as a function of the parameter $\chi$
obtained for magnetic fields of $\pm 1$~T
applied along the $y$-direction.
As addressed above this ratio yields the ratio between SIA and BIA strengths.
As an important result
Fig.~\ref{fig2}a shows that the SIA/BIA-ratio strongly depends on the doping position and, moreover,
changes its sign for $\chi \approx 0.1$.

The dependence of the both separate contributions $J_x$ and $J_y$ on
the parameter $\chi$ is shown in Fig.~\ref{fig2}b.
Here we normalize the data
by the free carrier concentration $n_s$
to enable a comparison of  BIA and SIA  in different samples.
For the Boltzmann distribution of carriers relevant to experiments at room temperature
the MPGE current is proportional to $n_s$ but independent of the electron mobility.~\cite{naturephysics06}
Thus, in order to compare the SIA and BIA contributions to MPGE in different samples
it is sufficient to normalize the photocurrents by the electron density $n_s$.

Figure~\ref{fig2}b  demonstrates that while the longitudinal current is 
almost independent of $\chi$ the transverse current strongly depends on $\chi$.
The results of Fig.~\ref{fig2}b  are in a full agreement with
 Eq.~(\ref{MPGE_second}) demonstrating that the
longitudinal current in this experimental geometry is solely due to BIA and the transverse
current is caused by SIA only.
In additional experiments we applied the magnetic field ${\bm B}$ along the $x$-direction.
We observed that while the  data on the transverse current remained the same
the longitudinal photocurrent inverted its sign in all samples. This observation is
also in agreement with  Eq.~(\ref{phen0}) yielding   for $\bm B \parallel x$ that
the photocurrent is given by $j_x= - \gamma^{\rm BIA}B_x|\bm E|^2$ and $j_y=\gamma^{\rm SIA}B_x|\bm E|^2$.

The fact that the longitudinal current in both experimental geometries
is independent of $\chi$ is expected for BIA-induced effects which are obviously insensitive
to the magnitude and the sign of $\chi$.
The transverse current, in contrast, is caused by SIA
and is very sensitive to the impurity potential. The variation of $\chi$ affects the degree of
asymmetry and even changes the sign of the transverse photocurrent due to SIA for $\chi \approx 0.1$.
Our results show that for $\chi < 0.1$ the asymmetry is dominated by the  potential of impurities
placed on the left (substrate) side of the QW. Figure~\ref{fig2}b demonstrates  that in order to obtain
a vanishing value of SIA (001)-oriented samples must be asymmetrically doped, so that the
$\delta$-doping layer on the substrate side is placed at a larger distance than that to the right of the QW ($l > r$).
This observation is attributed to the segregation of Si-impurities
during the structure growth and is in contrast to (110)-grown structures,
where symmetrical doping  results in a vanishing  SIA.~\cite{condmat110}
This essential difference stems from the difference in growth conditions.
Indeed, the growth temperature of high-quality (001)-oriented QWs is typically higher than 600$^\circ$C,
while (110)-structures are grown at  480$^\circ$C.~\cite{W2} The high growth
temperature of (001)-oriented  heterostructures  leads to substantial dopant
migration in the growth direction during the growth process  (segregation)
which affects mostly the substrate side of a QW and results in SIA in
symmetrically doped QWs.
In order to suppress the segregation we grew a symmetrically doped sample 
with reduced temperature during the $\delta$-doping 
($T_\delta = 490^\circ$C). We find in this case
that the MPGE current perpendicular to ${\bm B}$, which is caused solely by SIA,
is almost equal to zero. Thus, our result clearly demonstrates 
that the reduced growth temperature suppresses segregation, 
which, therefore, does not introduce additional structure inversion asymmetry.

The next important observation is that for $\chi = 0$ (sample 4) and $\chi \approx 0.17$
(samples 6 and 7)
$J_x/J_y  \approx \pm 1$ (see Fig.~\ref{fig2}a)
indicating that SIA and BIA have equal strengths.
In such samples the effects due to SIA and BIA, e.g. band spin-splitting,
cancel each other in either $[1 \bar{1}0]$ or [110] crystallographic
directions depending on the relative sign of SIA and BIA terms.~\cite{Averkiev08p114002}
For the transverse photocurrent caused by magnetic field applied along
one of $\langle 1 1 0 \rangle$-axes the phenomenological theory yields 
$j_{\perp} = (\gamma^{\rm SIA} \pm \gamma^{\rm BIA}) B |\bm E|^2$,
where the two signs correspond to the two magnetic field
orientations. At $\gamma^{\rm SIA} \approx \gamma^{\rm BIA}$ one
can expect substantial difference in transverse photocurrents  at
$\bm B $ applied along [1$\bar{1}$0] and [110]. Indeed, this
anisotropy was observed: e.g. in sample~4 characterized by close SIA
and BIA strengths we found that the transversal signal changes its magnitude
by a factor of six under rotation of in-plane magnetic field $\bm B$ by 90 degrees.

To conclude, we investigated SIA and BIA in (001)-oriented
GaAs/AlGaAs QWs at room temperature. The observed modulation of
the SIA/BIA-ratio by  a shift of the doping position demonstrates
that the impurity position plays an important role
in the SIA. We show that high growth temperatures
of high-quality (001)-oriented
GaAs QWs add an additional factor to the SIA due to impurity segregation.
Our measurements based on the MPGE
demonstrate that this method can be successfully applied to study
SIA and BIA in QW structures. We emphasize that  the method
can be applied even at room temperature, where
many other methods cannot be used.
By investigating
samples with a different $\delta$-doping profile
we obtained  QWs with almost equal magnitudes of Rashba and Dresselhaus
constants which should be characterized by extraordinary long spin relaxation times.

This work is supported by the DFG via programs
SPP~1285 and  SFB~689, RFBR,
Russian President grant, and ``Dynasty'' foundation --- ICFPM.

\section*{Appendix 1}

Polarized light gives rise to new roots of the MPGE due to the optical
excitation instead of relaxation which solely determines the
photocurrent generated by the  unpolarized radiation.~\cite{naturephysics06}
We used linearly polarized radiation with the
polarization vector of the incoming light rotated applying $\lambda/2$-plate.
By that the azimuth angle $\alpha$ is varied between $0^\circ$ and $180^\circ$
corresponding to all possible orientations of
the electric field vector in the ($xy$) plane.
For linearly polarized radiation at normal incidence and 
$\bm B$ parallel to $x$ or $y$ axes Eq.~\eqref{phen0} takes the form
\begin{eqnarray}
\label{100}
    j_\parallel &=& \pm D_1 B \pm D_3 B \cos{2\alpha} +  C_2 B \sin{2\alpha} , \\
    j_\perp &=&  C_1 B - C_2 B \cos{2\alpha} \pm D_2 B \sin{2\alpha}, \nonumber
\end{eqnarray}
where $\alpha$ is the angle between $\bm E$ and $\bm B$,
and $\pm$ correspond to $\bm B$ parallel to $x$ and $y$ axes, respectively.
The coefficients $C_i$ and $D_i$ are caused by SIA and BIA,
respectively. The corresponding contributions to Eq.~\eqref{100} reflect the different
symmetry of SIA and BIA: the SIA induced terms are invariant while the BIA
contributions change their signs under rotation of $\bm B$ and $\bm E$ by 90$^\circ$ around $z$ axis.
All photocurrent contributions given by Eq.~(\ref{100}) as well as polarization dependences have been detected
in our samples. The photocurrent contributions
and the ratio of the corresponding SIA- and BIA-related coefficients
behave upon a variation of the parameter $\chi$
in agreement with the data obtained using unpolarized radiation.
Our results show that in all investigated samples using of
polarized radiation does not lead to an increase of the method sensitivity.
In some other samples, however, MPGE due to the optical
excitation can be substantially larger than that due to relaxation.
Therefore the applicability of the method using polarized radiation demonstrated in our experiments
may be of importance.

\end{document}